\documentclass[%
amsmath,%
amssymb,%
preprint,
prl
]{revtex4}

\usepackage{dcolumn}
\usepackage{url}
\usepackage{pstricks}
\usepackage{graphicx}
\usepackage{bm}

\newcommand{\eqn}[1]{Eq.\,(\ref{#1})}
\newcommand{\eqns}[1]{Eqs.\,(\ref{#1})}
\newcommand{\noeqn}[1]{(\ref{#1})}
\newcommand{\fig}[1]{Fig.\,\ref{#1}}

\newcommand{\hide}[1]{}

\begin{document}

\title{%
  Role of Compaction Ratio in the Mathematical Model of Progressive Collapse
}
\author{Charles M. BECK}
%
%
\email{beck.charles_m@yahoo.com}
\date{14 April, 2008}
\begin{abstract}
  We derive a mathematical model of progressive collapse and examine
  role of compaction.
  Contrary to a previous result by Ba\v{z}ant and Verdure, J. Engr. Mech. ASCE 133 (2006) 308,
  we find that compaction slows down the avalanche by effectively increasing the
  resistive force.
  We compare currently available estimates of the resistive force, that of
  Ba\v{z}ant and Verdure (2006) corrected for compaction for World Trade Center (WTC) 2,
  and of Beck, www.arxiv.org:physics/0609105, for WTC 1 and 2.
  We concentrate on a damage wave propagating through the building
  before the avalanche that figures in both models:
  an implicit heat wave that reduces the resistive force of the building
  by 60\% in Ba\v{z}ant and Verdure (2006), or a wave of massive destruction that
  reduces the resistive force by 75\% in Beck (2006).
  We show that the avalanche cannot supply the energy to the
  heat wave as this increases the resistive force by
  two orders of magnitude.
  We thus reaffirm the conclusion of Beck (2006) that the avalanche is initiated in the
  wake of the damage wave.
\end{abstract}

\maketitle

\begin{figure}[htp]
  \centering
  \includegraphics[scale=0.3,clip]{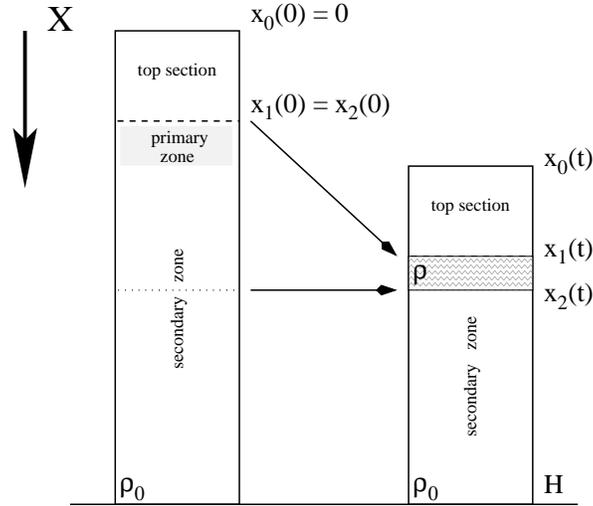}
  \caption{%
    \label{fig:avalanche}
    Propagation of an avalanche in a tall building of uniform density $\rho_0$.
    Extraneous factors cause initial weakening of the load bearing structure
    in the primary zone, leaving the building below (secondary zone) intact.
    The avalanche forms at the top of the primary zone, which propagates
    and compacts the building underneath from $\rho_0$
    to $\rho$, with $\kappa = \rho_0 / \rho \ll 1$ being a compaction ratio.
    The idealized scenario allows us to identify the following points:
    $x_0$, the top of the building; $x_1$, beginning of the compacted section of
    the avalanche; and $x_2$, the location of the avalanche front.
    Here it is implied that it is easier for the avalanche to drop (``crush down'')
    then to stop and compact the top section instead (``crush up'').
  }
\end{figure}

Ba\v{z}ant and Verdure\cite{bazant2006} proposed the following mathematical
model to describe the progressive collapse in a tall building of a
homogeneous longitudinal density $\rho_0$,
\begin{equation}
  \label{eq:bazant:1}
  \rho_0 \, \left(1 - \kappa \right) \,
  \frac{d}{dt} \left( x_2 \, \dot x_2 \right) = {R}(x_2) + \rho_0 \, g \, x_2,
\end{equation}
where $x_2$ is the position of the avalanche front, $\rho_0 = M/H$ with $M$ the
total mass and $H$ the total height of the building and $g$ is the gravity,
while $R = R(x_2)$ is a local resistive force.
Here, the dot above the quantity indicates its differentiation with respect to time.
The compaction ratio $\kappa$ is defined as
\begin{equation}
  \label{eq:kappa}
  \kappa = \frac{\rho_0}{\rho},
\end{equation}
where $\rho$ is a density of the ``compacted'' section of the building, cf. \fig{fig:avalanche}.

We reexamine the steps that lead to \eqn{eq:bazant:1} from the point of
view of classical mechanics.
As can be seen from \fig{fig:avalanche} we can choose between two
generalized coordinates.
The first choice is $x_1 = x_{1}(t)$ which for $t>0$ well describes the
motion of the avalanche.
The second choice is $x_2 = x_2(t)$, which represents the position of
the avalanche front - an idealized point-like boundary between the stationary and the
moving part of the building, at which the compaction takes place.
Motion of the avalanche front is more complex than the motion of the avalanche
as it combines the motion of the avalanche with its spatial growth due to
non-zero compaction ratio.
While care must be exercised when deriving an equation of motion for each of them,
the final result may not depend on the choice of generalized coordinate.

For simplicity, we assume that the total energy and the total mass in the
system building-avalanche is conserved.
We note that in that case we obtain the fastest avalanche,
as then there are no conversion losses of the potential energy of the building
into the kinetic and then into the crushing energy of the avalanche.
Also, the conservation of energy allows us to use Lagrangian formalism to derive
the equation of motion.

First, we state the two constraints of descent,
\begin{subequations}
  \label{eq:constraint}
  \begin{equation}
    \label{eq:constraint:1}
    x_1 - x_0 =  h,
  \end{equation}
  \begin{equation}
    \label{eq:constraint:2}
    \rho_0 \, H = \rho_0 \, \left( x_1-x_0 \right)
    + \rho \, \left( x_2-x_1 \right)
    + \rho_0  \, \left( H - x_2 \right).
  \end{equation}
\end{subequations}
Here, the top section stretches from $x_0 = x_0(t)$ down to $x_1 = x_1(t)$,
while the compacted building occupies from $x_1$ down to $x_2=x_2(t)$.
The point $x_2$ is the avalanche front.
With \eqn{eq:constraint:1} we state that the length of the top section
$h$ does not change in descent,
while with \eqn{eq:constraint:2} we express the conservation of the mass
of the building.
Differentiation of \eqn{eq:constraint} with respect to time
yields dynamical constraints, $\dot x_1 = \dot x_0$ and
$\dot x_1 = (1-\kappa) \, \dot x_2$.

Second, we find kinetic, potential and latent energy necessary
for the Lagrangian formulation.
The kinetic energy $K$ is given by $K = \frac 1 2 \, \int dx \, \rho(x) \, v^2(x)$,
where the velocity distribution in the avalanche is
$v(x) = \dot x_1$ for $x \in \left[x_0, x_2 \right>$.
The kinetic energy of the avalanche is thus
\begin{equation}
  \label{eq:ke}
  K = \frac{1}{2} \rho_0 \, x_2 \, \dot x_1^2
    = \frac{1}{2} \rho_0 (1-\kappa)^2\, x_2 \, \dot x_2^2.
\end{equation}
The potential energy $U$ of the whole building is $U = -\int dx \,\rho(x) \, x \, g$,
yielding
\begin{equation}
  \label{eq:pe}
  U = -\frac 1 2 \, g \, \rho_0 \,\left(
    H^2 + (1-\kappa) \, ( x_2^2 - h^2 )
    \right).
\end{equation}
The latent energy $L = L(x_2)=-\int^{x_2}_H\,dx'\,R(x')$, produces
the resistive force of the building,
$R(x_2) = - \partial{L(x_2)}/\partial x_2$.

Given a Lagrangian ${\cal L} = K - U - L$,
which is a function of a generalized coordinate $x$ and its generalized
velocity $\dot x$,
the equation of motion follows from
$\frac {d}{dt} \, \partial{\cal L}/\partial \dot x = \partial{\cal L}/\partial x$.
We recall that we have two choices for the generalized coordinate:
$x \equiv x_1$ for the motion of avalanche, or $x \equiv x_2$ for the avalanche front.
If $x_2$ is chosen as a generalized coordinate, we simply obtain,
\begin{equation}
  \label{eq:eom:x2}
  \rho_0 \, (1-\kappa)^2 \frac {d}{dt} \left( x_2 \, \dot x_2 \right)
  =
  R(x_2)
  + \rho_0 \, g \, (1 - \kappa) \, x_2
  + \frac{\eta \, \rho_0}{2} \, (1 - \kappa)^2 \, \dot x_2^2.
\end{equation}
With $x_1$ as a generalized coordinate we note that
$\partial /\partial x_1 = (1-\kappa)^{-1} \partial/\partial x_2$,
yielding
\begin{equation}
  \label{eq:eom:x1}
  \rho_0 \, (1-\kappa) \frac {d}{dt} \left( x_2 \, \dot x_2 \right)
  =
  \frac{R(x_2)}{1-\kappa}
  + \rho_0 \, g \, x_2
  + \frac{\eta \, \rho_0}{2} \, (1 - \kappa) \, \dot x_2^2.
\end{equation}
As expected, the two equations of motion are identical.
In \eqns{eq:eom:x2} and \noeqn{eq:eom:x1} we introduced an additional
parameter $\eta$ which may take values
1, if the total energy is conserved, or 0, if this is not the case.
A distinction between the two cases is discussed in~\cite{pesce2003}.
Comparison between \eqn{eq:eom:x1} and \eqn{eq:bazant:1} of Ba\v{z}ant
and Verdure's shows that due to non-zero compaction the avalanche
propagates through the building faster than it travels, which leads to
an amplification of the building's resistive force by the factor $(1-\kappa)^{-1} > 1$.
In other words, the avalanche front in Ba\v{z}ant and Verdure's model, \eqn{eq:bazant:1},
is faster than the one proposed in \eqn{eq:eom:x1}.
\eqn{eq:eom:x1} as a proposed correction to \eqn{eq:bazant:1} is a major
result of this technical note.
We next discuss the estimates for the resistive force $R$ that can be found
in the literature.

Ba\v{z}ant and Verdure~\cite{bazant2006} analyzed collapse of
World Trade Center 2 in terms of a model~\noeqn{eq:bazant:1}.
They assumed that $R$ is a constant throughout the building's
primary and secondary zone, where $R = \Delta L / \Delta H$.
For a crushing energy they made an educated guess, $\Delta L = 2.4$~GNm,
with the floor height being $\Delta H = 3.7$~m.
Considering the total mass of the building to be $M=3.2\cdot10^{8}$~kg,
this yields for the resistive force $R/(M\,g) = r =  0.2$, as their initial
estimate.
They noted that in order for the avalanche to reach the ground level
in $_2T \simeq 10.8$~s~\footnote{%
  $_2T \simeq 10.8$~s is an unofficial estimate of the duration of collapse of
  World Trade Center 2.
}
they had to use $R/2$ instead of $R$.
The 50\% reduction, they argued, came from heat (p.15, top paragraph of the on-line
edition of~\cite{bazant2006}).
That is, an assumption built in their model is that the avalanche
pushes a heat wave in front of itself which reduces the strength of the building by 50\%.
Using their value for compaction $\kappa \simeq 0.2$ and
the corrected equation of motion proposed here, requires $R$ to be reduced
by 60\% instead.\\
Here the following comment is in place.
If the avalanche were supplying energy to the heat wave
then $R$ in~\eqn{eq:bazant:1} splits in two components.
First, original $R$ decreases, say, by 60\% as discussed earlier.
However, an additional resistive force $(R)_{heat}$ appears which describes
the rate of transfer of energy from the avalanche to the heat wave per unit length.
A simple estimate shows~\footnote{%
  This estimate for the magnitude of the resistive force due to heat dissipation
  follows from
  $(R)_{heat} = dW_{heat}/dx_2$,
  with $W_{heat} = \sigma\,(m(x_2)-m(x_1))\,C\,\Delta T$,
  and $m(x) = \rho_0 \, x$ assuming the uniform distribution of mass.
  Taking the heat capacity of the steel $C \simeq 486$~J/(kg\,K),
  $H=417$~m for the height of the building,
  $\Delta T=300$~K for the increase in the steel temperature at which
  steel strength is reduced by 50\%,
  and $0.5\le\sigma \le 0.9$ for the ratio of the mass of steel to the total
  mass of the building.
  This yields $(R)_{heat}/(M\,g) = \sigma \, C \, \Delta T / (H\, g) = 17.8$,
  which is by two orders of magnitude greater than $r\sim0.2$,
  the apparent resistive force of the buildings.
} that $(R)_{heat}/(M\,g) \sim 20 \gg R$.
It is thus obvious that the 60\%-strength-reducing heat wave could not have
been created or maintained by the avalanche.
On the contrary, if the avalanche were in fact heating the steel, then this acted as a
resistive force comparable to  $R$.
E.g., the avalanche warming the steel by $\Delta T = 10$~K yields
$(R)_{heat}/(M\,g) =r  \sim 0.1$ which is comparable to the observed resistive force.
We do mention, however, that the authoritative document on collapse of
World Trade Center 1 and 2, the NIST report~\cite{nist:2005} explicitely
states that no elevated temperatures were observed in the secondary zone.

Rather than guessing, we proposed a procedure for estimating $R$~\cite{beck2006},
where one first finds the ultimate yield strength $Y$  of the vertical columns using their
specifications, following which the resistive force $R$ is estimated from a simple
linear model, $R = \epsilon \cdot Y$, with $\epsilon = 0.25$
being the ultimate yield strain of structural steel under compression.
Applying this to WTC 1 and 2 led to an initial estimate of
$R(x_2)/(M\,g) = r + s \cdot (x_2/H)$, with $r\simeq0.2$ and $s\simeq0.7$.
We analyzed the collapse of World Trade Centers 1 and 2 where we divided the
building into the primary and the secondary zone where we allowed $R_I$ in the primary
zone to be considerably smaller than $R_{II}$ in the secondary zone,
$R_I = 1/4 \cdot R_{II}$ for WTC 1 and $R_{I} = 3/8 \cdot R_{II}$ for WTC 2,
while we neglected the compaction altogether.
We did this subdivision after the statement from the NIST report that
the damage to the buildings was concentrated in their primary zones
while leaving the secondary zones intact.
We found that to reach the collapse time $_2T$ the initial estimate of $R$ had to
be reduced by 75\%.
In our report we dubbed the 75\%-strength-reducing wave that preceeded
the avalanche the wave of massive destruction (WMD).
As the avalanche was not supplying the energy to the WMD, and the WMD propagated
before the avalanche, we concluded that the WMD caused the avalanche.

Currently, we can only speculate about the source of the 60-75\%-strength-reducing
wave and its coupling to the avalanche.
However, a piece of information that would provide an important insight is
the descent curve, which describes position as a function of time of some visible part of
the building, say its top, $x_0 = x_0(t)$.
Once the descent curve is known it is the acceleration,
$\ddot x_0 = \ddot x_0(t)$, that can be directly connected to $R$ through a
mathematical model.
In fact, in~\cite{beck2008} we examine the descent curve available for WTC 7 in terms of
a corrected model of ``crush-up'' mode of progressive collapse and identify the
phases of descent.

Lastly, as the collapses of World Trade Centers resemble controlled
demolition it would be instructive to apply the methods discussed in
this and other articles to other buildings that are known to have
been destroyed in controlled fashion.



\section{Figures}

\end{document}